%% file: esrel2025.tex
\pdfoutput=1

\documentclass[twocolumn]{rps-esrel2025}

\def\papername{\jobname}

\usepackage{makecell}   %Make cell for table
\usepackage{mwe}        %wrap figure
\usepackage{subcaption} %subfigure environment
\usepackage{multicol}
\usepackage{multirow}
\usepackage{wrapfig, booktabs}
\usepackage{amsmath}    %for AFDT math definition
\usepackage{amssymb}
\usepackage[normalem]{ulem}

\usepackage{pifont}
\usepackage{bbding}

\usepackage{soul} %highligh \hl
\usepackage{mathrsfs} %mathSCR

\usepackage{graphicx}      
\usepackage[dvipsnames]{xcolor}
\definecolor{component}{RGB}{34,94,168} % 
\definecolor{attack}{RGB}{252, 78, 42} % 
\definecolor{defense}{RGB}{51, 170, 51} % 51, 255, 51
\definecolor{gate}{RGB}{136,65,157} % 

\usepackage{pgfplots}   

\usepackage{comment}
\newif\ifsubmit

\usepackage{natbib}

\usepackage{comment}

\input{macros}

\begin{document}

\markboth{Soltani et al.}{Safety and Security Risk Mitigation in Satellite Missions via Attack-Fault-Defense Trees}

%%%%%%%%%%%%%%%%%%%%%%%%% Plase keep this command for single column for abstract section.
\twocolumn[
%%%%%%%%%%%%%%%%%%%%%%%%%

\title{Safety and Security Risk Mitigation in Satellite Missions via Attack-Fault-Defense Trees}

\author{Reza Soltani}

\address{Formal Methods and Tools group, University of Twente, Enschede, the Netherlands. \email{r.soltani@utwente.nl}}

\author{Pablo Diale}

\address{Ascentio Technologies S.A., Cordoba, Argentina. \email{pdiale@ascentio.com.ar}}

\author{Milan Lopuhaä-Zwakenberg}

\address{Formal Methods and Tools group, University of Twente, Enschede, the Netherlands. \email{m.a.lopuhaa@utwente.nl}}

\author{Mariëlle Stoelinga}

\address{Formal Methods and Tools group, University of Twente, Enschede, the Netherlands;} 
\address{Radboud University, Nijmegen, the Netherlands. \email{m.i.a.stoelinga@utwente.nl}}

\begin{abstract} 
Cyber-physical systems, such as self-driving cars or digitized electrical grids, often involve complex interactions between security, safety, and defense. Proper risk management strategies must account for these three critical domains and their interaction because the failure to address one domain can exacerbate risks in the others, leading to cascading effects that compromise the overall system resilience. This work presents a case study from Ascentio Technologies, a mission-critical system company in Argentina specializing in aerospace, where the interplay between safety, security, and defenses is critical for ensuring the resilience and reliability of their systems. The main focus will be on the Ground Segment for the satellite project currently developed by the company. Analyzing safety, security, and defense mechanisms together in the Ground Segment of a satellite project is crucial because these domains are deeply interconnected—for instance, a security breach could disable critical safety functions, or a safety failure could create opportunities for attackers to exploit vulnerabilities, amplifying the risks to the entire system. This paper showcases the application of the Attack-Fault-Defense Tree (AFDT) framework, which integrates attack trees, fault trees, and defense mechanisms into a unified model. AFDT provides an intuitive visual language that facilitates interdisciplinary collaboration, enabling experts from various fields to better assess system vulnerabilities and defenses. By applying AFDT to the Ground Segment of the satellite project, we demonstrate how qualitative analyses can be performed to identify weaknesses and enhance the overall system’s security and safety. This case highlights the importance of jointly analyzing attacks, faults, and defenses to improve resilience in complex cyber-physical environments.
\end{abstract}


%%%%%%%%%%%%%%%%%%%%%%%%% Please keep this closing bracket to complete the single column format for abstract.
]
%%%%%%%%%%%%%%%%%%%%%%%%%

\input{Sections/1-Intro}

\input{Sections/2-RelatedWork}
\input{Sections/3-AFDT}
\input{Sections/4-AscentioCS}
\input{Sections/5-Conclusion}

\begin{acknowledgement}
The work reported in this paper was partially funded by the NWO under grants PrimaVera (NWA. 1160. 18. 238),
ZORRO (KICH1. ST02. 21. 003), Marie Skłodowska-Curie grant MISSION (101008233), ERC under grants CAESAR (864075), and RUBICON (101187945).
\end{acknowledgement}

\bibliographystyle{chicago}
\bibliography{esrel2025}

\end{document}

%% file: macros.tex
\newcommand{\colorpar}[3]{\colorbox{#1}{\parbox{#2}{#3}}}
\newcommand{\marginremark}[3]{\marginpar{\colorpar{#2}{4em}{\color{#1}#3}}}

\ifsubmit
  \newcommand{\rs}[1]{}
  \newcommand{\ms}[1]{}
  \newcommand{\mlz}[1]{}
\else
  \newcommand{\rs}[1]{\marginremark{black}{yellow}{\tiny{[RS]~ #1}}}
  \newcommand{\ms}[1]{\marginremark{white}{blue}{\tiny{[MS]~#1}}}
  \newcommand{\mlz}[1]{\marginremark{purple}{white}{\tiny{[MLZ]~ #1}}}
\fi

\newcommand{\xmark}{\ding{55}}%%% Comments

%% file: Sections/1-Intro.tex
\section{Introduction}

In recent years, the satellite industry has witnessed a paradigm shift with the emergence of Ground Segment as a Service (GSaaS), which enables satellite operators to outsource ground segment operations, including data processing, mission control, and communication infrastructure, to specialized service providers. This innovative model enables satellite operators to access and manage ground segment infrastructure on a pay-as-you-go basis, eliminating the need for substantial capital investments in proprietary ground systems. By leveraging GSaaS, operators can expedite mission deployment, reduce operational costs, and concentrate on core business activities such as data provision and analysis. 

However, the adoption of GSaaS introduces new challenges, particularly in ensuring the safety and security of satellite operations. The shared nature of GSaaS infrastructure necessitates robust mechanisms to protect against potential threats and system failures. 

In cyber-physical systems, safety and security are frequently investigated separately in different studies. Nevertheless, there is a strong interdependency between them \citep{survey}. In the complex safety and security interplay that involves trade-offs, measures that improve security may weaken safety or vice versa. 
We need to increase the resilience of critical infrastructures such as GSaaS for satellite operators not only to accidental failures that may come from many high-tech components but also to (cyber)attacks by malicious actors. To achieve high resilience against such risks, we may consider using countermeasures against safety and security risks. However, experts from several fields must collaborate on such implementation, which leads to the need to have a common framework for assessing the safety, security, and impact of countermeasures. 

Tree-based models are ubiquitous in both safety and security risk assessment. Fault Trees (FTs) \citep{FT1,FT2} are introduced for safety, and Attack Trees (ATs) \citep{AT} for security. These are often used frameworks for allowing communication across disciplines. 
To capture the wide range of risks and associated countermeasure strategies, more comprehensive models are required, as FTs and ATs only address safety and security, respectively. 
There are frameworks for joint analysis like Attack-Defense Trees (ADTs), which model security risks and countermeasures to mitigate them, and Attack-Fault Trees (AFTs), which represent joint safety-security risks. However, none of these models has the expressive power to model the interaction between safety, security, and defense that is inherent to critical infrastructures such as GSaaS, particularly in scenarios where system failures, cyberattacks, and defensive measures interact in complex and unpredictable ways.

To address these concerns, we introduced a new framework, namely Attack-Fault-Defense Trees (AFDTs), that captures all safety, security, and defense domains in a single framework \citep{AFDT}. 
In our previous work \citep{AFDT}, we presented the mathematical definition of AFDTs and their structure-function along with the semantics and cut set metrics. In addition, we provided a case study of a power grid to showcase the application of our framework \citep{AFDT2}. 
While that study primarily emphasized safety aspects, aligning with the domain and requirements of that specific application, it also had a limited number of defenses.
We address these limitations by presenting a new case study and applying AFDTs to a GSaaS environment, which is inherently more security-dominant, that works better for AFDTs with many defenses. 
We also perform qualitative and quantitative risk analyses, showcasing the scalability and applicability of AFDTs in enhancing the resilience of satellite ground segment services.

%% file: Sections/2-RelatedWork.tex
\section{Related Work}

In the safety and security domain, tree-based formalisms form the majority of formalisms that capture the interplay between security and safety \citep{survey}.
Attack Trees (ATs) \citep{AT} deal with system attacks, while Fault Trees (FTs) \citep{FT1,FT2} were made to handle system failures. 
In a survey of models for safety-security co-analysis, \cite{survey} discovered that there is no model that precisely represents safety-security interactions.
Instead, a variety of methods are used to combine constructs from frameworks that only concentrate on security or safety.
Metrics are not distinct; not one is uniquely designed with safety/security interactions in mind.
Furthermore, there is a shortage of large-scale case studies, and current formalisms only model dependencies in small- and medium-sized case studies.

\begin{table*}
\centering
\scriptsize % Reduce font size
\caption{Related works comparison to the proposed approach}
\label{related_works}
\begin{tabular}{cc|ccccc}
       %\hline
     \multicolumn{2}{c|}{Model} & Attack & Failure & \thead{Defense/\\Countermeasure} & \thead{Qualitative/Quantitative\\analysis} & Case study\\
       \hline
     FT & \cite{DFT} &  & \Checkmark &  & \Checkmark &  \\
         \hline 
     \multirow[c]{4}{*}[0in]{ADT}& \cite{Kordy-game} & \Checkmark &  & \Checkmark &  &  \\
         %\hline
     & \cite{ADT1} & \Checkmark &  & \Checkmark & \Checkmark &  \\
         %\hline
     & \cite{ACT} & \Checkmark &  & \Checkmark & \Checkmark & \Checkmark \\
         %\hline   
     & \cite{CSF} & \Checkmark &  & \Checkmark & \Checkmark & \Checkmark \\
         \hline
     AT & \cite{CSF-29} & \Checkmark &  & \Checkmark & \Checkmark & \Checkmark \\
         \hline
     \textbf{AFDT} & \cite{AFDT} & \Checkmark & \Checkmark & \Checkmark & \Checkmark & \Checkmark \\
         %\hline
\end{tabular}
\end{table*}

Due to differences in how they are used, FTs and ATs are extended either with additional gates and system recovery \citep{DFT,ACT} or defenses \citep{ADT1,ADT2}.
\cite{ADT1} define attack–defense trees as attack trees with defenses in the form of countermeasures.
\cite{CSF} look into the most effective countermeasures for ADTs. The techniques used by \cite{CSF-29} to determine the optimal countermeasures in attack graphs, an alternative risk model for security, rely on the activities of each defender, which affects the probability that an attack will be successful.
\cite{FACT} developed a failure-attack-countermeasure graph architecture to align safety and security during the early stages of the development of cyber-physical systems. 
Safety, security, and countermeasures are only included in the graph in the early stages of development. 
The paper does not include a semantic or qualitative analysis of the graph.

We compared related work to the AFDT approach in Table 1. As can be seen, AFDT is unique in combining failures, attacks, and defenses. 
In our previous work \citep{AFDT2}, we showcased the application of the AFDT framework through a safety-dominant case study, which primarily focused on safety with a limited number of defenses. In this paper, we extend our approach to a more security-dominant case study, applying AFDT to a GSaaS environment for satellite operations, which incorporates a greater number of defenses to address its distinct challenges.

\begin{figure*}
     \centering
     \begin{subfigure}[b]{0.075\textwidth}
         \centering
         \includegraphics[width=\textwidth]{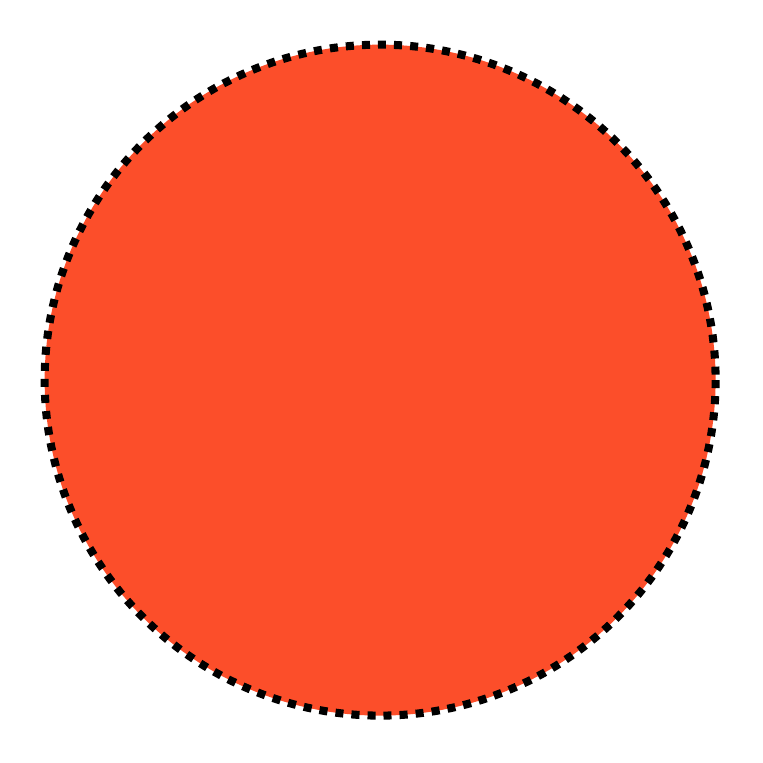}
         \caption{BAS}
         \label{bas}
     \end{subfigure}
     \hfill
     \begin{subfigure}[b]{0.075\textwidth}
         \centering
         \includegraphics[width=\textwidth]{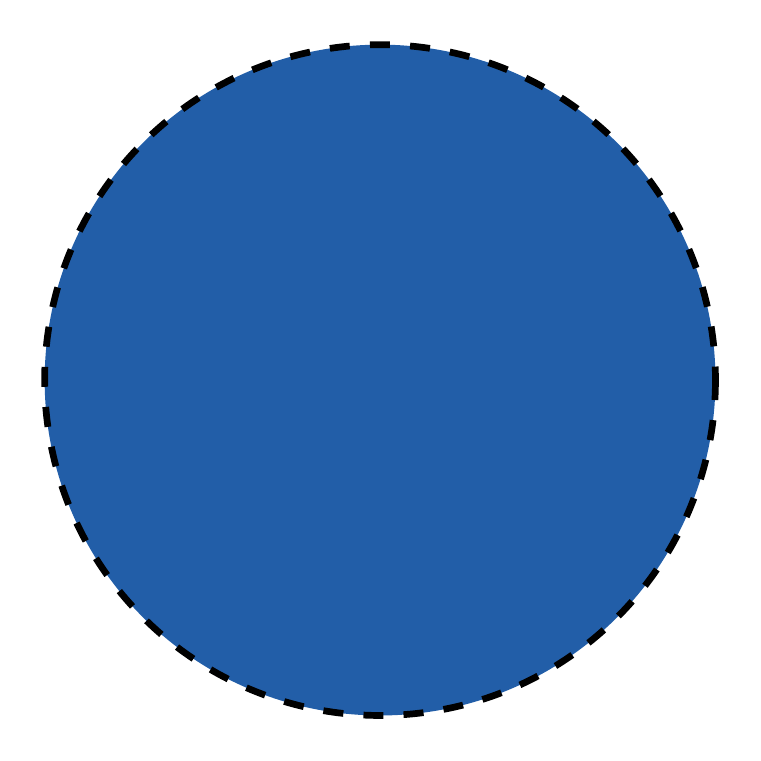}
         \caption{BCF}
         \label{bcf}
     \end{subfigure}
     \hfill
     \begin{subfigure}[b]{0.075\textwidth}
         \centering
         \includegraphics[width=\textwidth]{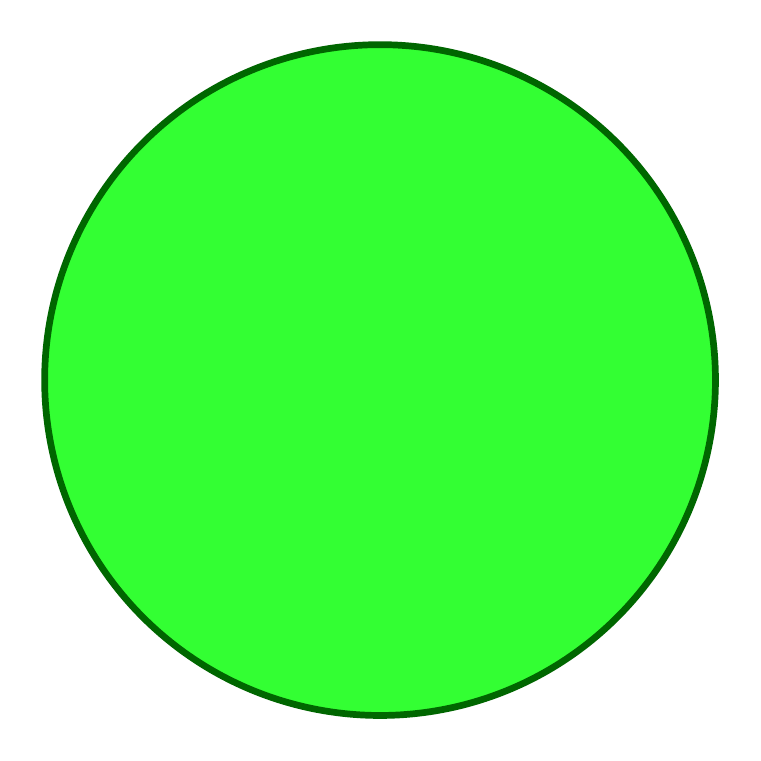}
         \caption{BDS}
         \label{bds}
     \end{subfigure}
     \hfill
     \begin{subfigure}[b]{0.08\textwidth}
         \centering
         \includegraphics[width=\textwidth]{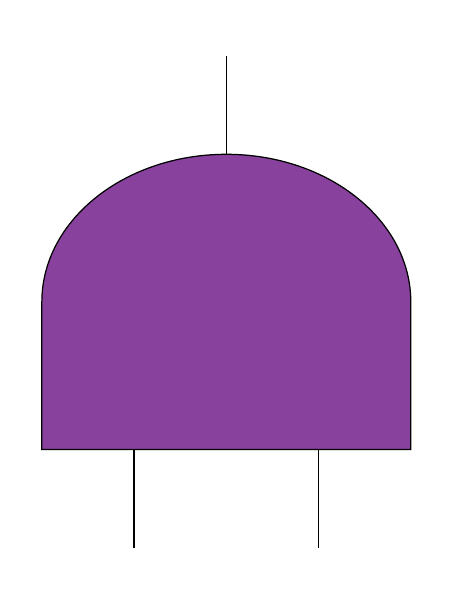}
         \caption{AND}
         \label{and}
     \end{subfigure}
     \hfill
     \begin{subfigure}[b]{0.08\textwidth}
         \centering
         \includegraphics[width=\textwidth]{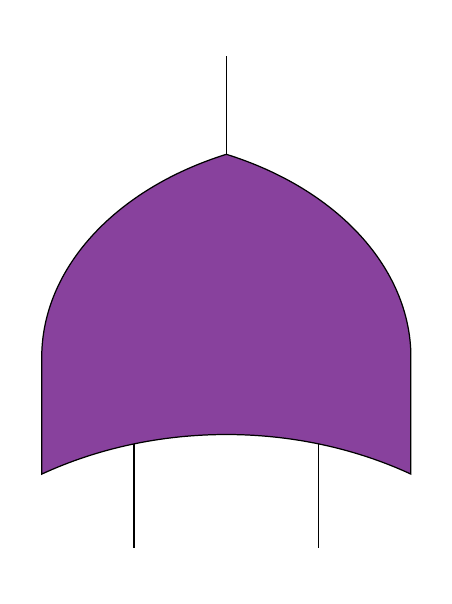}
         \caption{OR}
         \label{or}
     \end{subfigure}
     \hfill
     \begin{subfigure}[b]{0.12\textwidth}
         \centering
         \includegraphics[width=\textwidth]{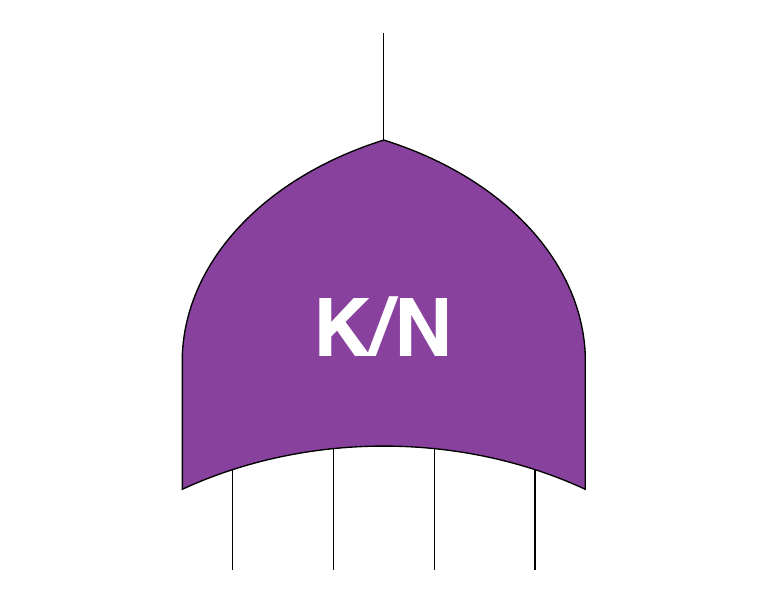}
         \caption{VOT(k/n)}
         \label{vot}
     \end{subfigure}
     \hfill
     \begin{subfigure}[b]{0.09\textwidth}
         \centering
         \includegraphics[width=\textwidth]{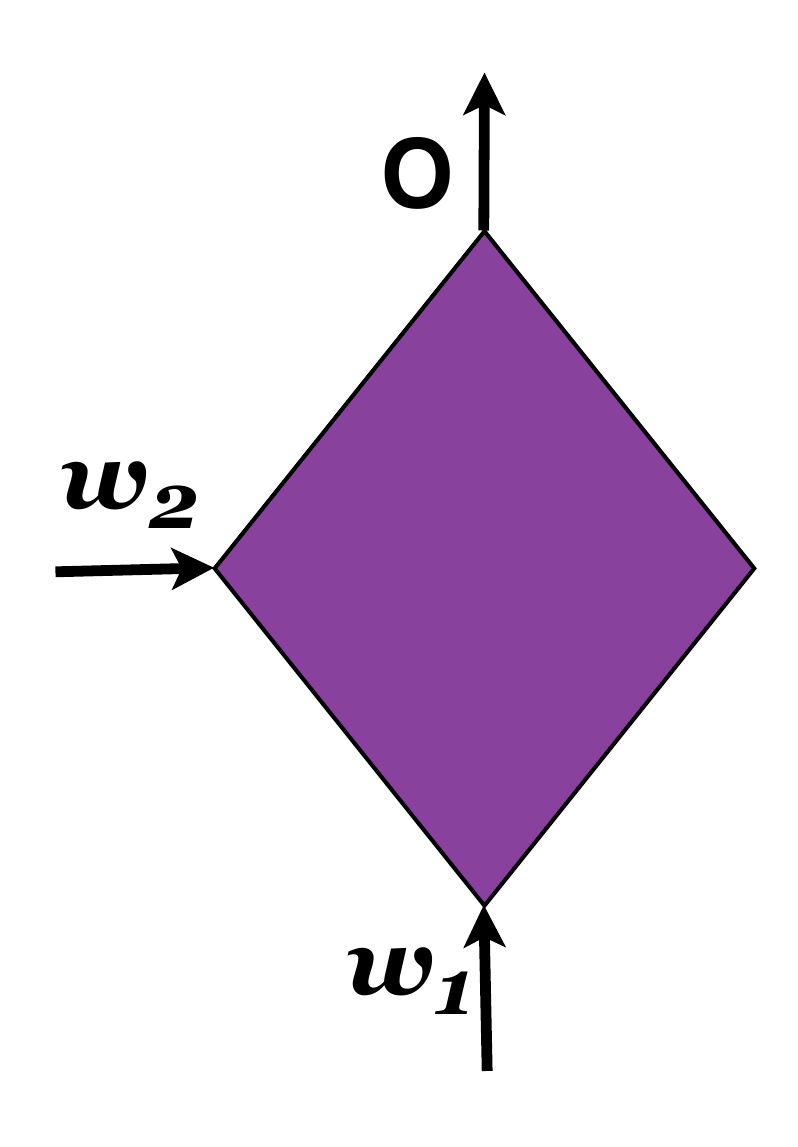}
         \caption{INH}
         \label{defense}
     \end{subfigure}
        \caption{Attack-Fault-Defense tree elements}
        \label{afdt_element}
\end{figure*}

%% file: Sections/3-AFDT.tex
\section{Background: AFDT}
\label{background}

\paragraph{AFDT.}The various elements of AFDT \citep{AFDT} are illustrated in Figure \ref{afdt_element}, with Figs. \ref{bas}--\ref{bds} depicting the various leaf types as circular nodes, such as the Basic Attack Step (BAS), Basic Component Failure (BCF), and Basic Defense Step (BDS). The interdependence between leaves happens through logical symbols called gates. 
These gates mirror (illustrate) discrete events and have inputs from other gates or individual leaves. 
We considered the gate types \emph{AND} (Fig. \ref{and}), \emph{OR} (Fig. \ref{or}), \emph{VOT(k/n)} (voting, Fig. \ref{vot}), and \emph{INH} (inhibition, Fig. \ref{defense}) in the analysis, that activate when resp. all, one, and $k$ of the $N$ inputs are activated. The integrity of the system is compromised when the TLE is activated.

\begin{figure}
     \centering
        \includegraphics[width=0.4\textwidth]{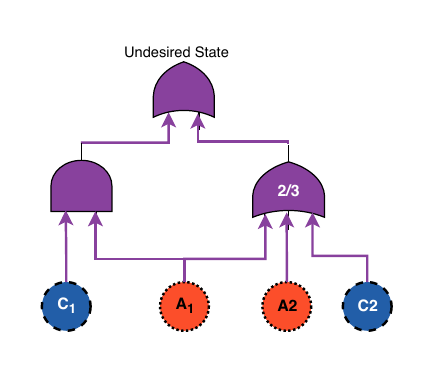}
        \caption{An AFT example. The system fails if either $\textcolor{component}{C_1}$ and $\textcolor{attack}{A_1}$ both occur, or if at least 2 of $\textcolor{attack}{A_1},\textcolor{attack}{A_2},\textcolor{component}{C_2}$ occur.}
        \label{toy1}
\end{figure}

When both BCFs and BASs are present, the model is an Attack-Fault Tree (AFT). AFT models both safety and security and their interaction. An example of an AFT is shown in Fig. \ref{toy1} with two \textcolor{component}{BCFs} and two \textcolor{attack}{BASs}. 
The TLE is triggered when both the component failure \textcolor{component}{$C_1$} and the attack \textcolor{attack}{$A_1$} occur, or when the failure \textcolor{component}{$C_2$} coincides with the attacks \textcolor{attack}{$A_1$} and \textcolor{attack}{$A_2$}, under the condition that at least two of these three events must occur.

\begin{figure}
     \centering
        \includegraphics[width=0.35\textwidth]{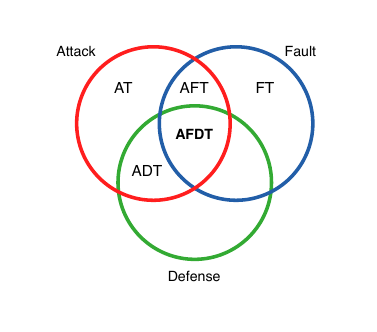}
        \caption{Venn diagram representing the tree-based formalisms discussed in this paper.}
        \label{Venn}
\end{figure}

AFDTs, introduced in our previous work \citep{AFDT}, extend AFTs by incorporating defenses to prevent the propagation of safety and security risks. This formalism unifies the strengths of AFT and ADT, allowing for a comprehensive analysis of faults, attacks, and defenses within a single framework. See Fig.~\ref{Venn} for a visual comparison of tree-based formalisms.

\begin{figure}
     \centering
        \includegraphics[width=0.5\textwidth]{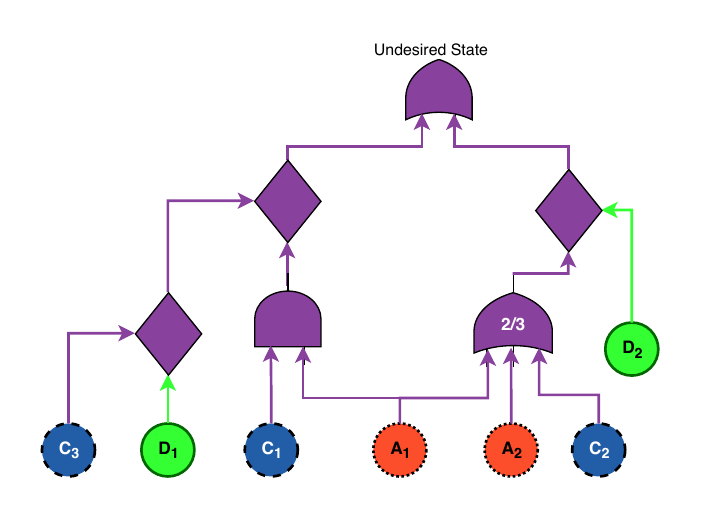}
        \caption{An exemplary depiction of AFDT, which extends Fig. \ref{toy1} by defenses $D_1$ and $D_2$}
        \label{toy2}
\end{figure}

Compared to AFT, AFDT introduces two key elements: the Basic Defense Step (BDS), representing atomic actions by the defender to enhance system resilience, and the \emph{INH}-gate, which models countermeasures that prevent specific events from propagating when triggered. Figure~\ref{toy2} illustrates how defenses (e.g., \textcolor{defense}{$D_1$} and \textcolor{defense}{$D_2$}) are added to the AFT in Figure~\ref{toy1}, showcasing their role in mitigating risks such as \textcolor{component}{$C_1$} with \textcolor{attack}{$A_1$}, and the joint propagation of \textcolor{attack}{$A_1$}, \textcolor{attack}{$A_2$}, and \textcolor{component}{$C_2$}. Disabling relations, modeled by \emph{INH}-gates, capture scenarios where defenses can fail due to components like \textcolor{component}{$C_3$}. Full formal details of AFDTs are available in \cite{AFDT}.

\begin{table}[]
    \centering
    \scriptsize % Reduce font size
    \caption{MCS analysis of the toy example AFDT} 
    \begin{tabular}{|c|c|c|c|}
    \hline
        No defense & \{\textcolor{defense}{$D_1$}\} & \{\textcolor{defense}{$D_2$}\} & \{\textcolor{defense}{$D_1$}, \textcolor{defense}{$D_2$}\} \\
    \hline
    \{\textcolor{component}{$C_1$}, \textcolor{attack}{$A_1$}\} & \{\textcolor{component}{$C_3$}, \textcolor{component}{$C_1$}, \textcolor{attack}{$A_1$}\} & \{\textcolor{component}{$C_1$}, \textcolor{attack}{$A_1$}\} & \{\textcolor{component}{$C_3$}, \textcolor{component}{$C_1$}, \textcolor{attack}{$A_1$}\}\\
         
    \{\textcolor{attack}{$A_1$}, \textcolor{attack}{$A_2$}\} & \{\textcolor{attack}{$A_1$}, \textcolor{attack}{$A_2$}\} & \xmark & \xmark\\
         
    \{\textcolor{attack}{$A_1$}, \textcolor{component}{$C_2$}\} & \{\textcolor{attack}{$A_1$}, \textcolor{component}{$C_2$}\} & \xmark & \xmark\\
    
    \{\textcolor{attack}{$A_2$}, \textcolor{component}{$C_2$}\} & \{\textcolor{attack}{$A_2$}, \textcolor{component}{$C_2$}\} & \xmark & \xmark\\
         \hline   
    \end{tabular}
    \label{tab:MCS}
\end{table}

\vspace{-0.5cm}
\paragraph{Qualitative analysis.} AFDTs provide a complete, system-wide view of how safety and security hazards might merge to cause system-level failures. It depicts how attacks and failures propagate to a higher level, resulting in the failure of a top-level event (TLE), and how defenses prevent this propagation. 
Finding the Minimal Cut Sets (MCSs) for AFTs is one of the most used methods of qualitative risk analysis. 
An MCS is a minimum collection of BCFs/BASs that, when added together, activate the TLE; if any of these elements are removed, the TLE becomes inactive.
Consider the AFT in Fig.~\ref{toy1}. It contains  four MCSs: \{\textcolor{component}{$C_1$}, \textcolor{attack}{$A_1$}\}, \{\textcolor{attack}{$A_1$}, \textcolor{attack}{$A_2$}\}, \{\textcolor{attack}{$A_1$}, \textcolor{component}{$C_2$}\}, and \{\textcolor{attack}{$A_2$}, \textcolor{component}{$C_2$}\}.
Any of these MCSs gives information about the system's vulnerability and also depicts the most concise path that caused the activation (failure) of the TLE. 
Each MCS has a finite number of elements.

An AFDT's identification of MCSs is different from an AFT's because the former has protections that can prevent an attack from propagating. 
We have four sets of MCSs in the AFDT of Fig. \ref{toy2} since there are two BDSs.
An overview of the MCSs corresponding to each defense activation for AFDT of Fig. \ref{toy2} is provided in Table \ref{tab:MCS}.
Table \ref{tab:MCS} indicates that the defense \textcolor{defense}{$D_2$} activation results in the removal of three MCSs, specifying its effectiveness. 
Defense \textcolor{defense}{$D_1$} increases the element count of one MCS but has no effect on another. 
System dependability is increased when both defenses are activated at the same time, creating a three-element MCS. 
It should be noted that increasing the size of an MCS, as well as deleting it entirely, increases the reliability of a system.
MCS analysis in AFDT provides an overview of each defense's influence and relationship to system reliability.
We can analyze MCS quantitively when data is available by considering the defense’s effect on MCS parameters, such as probability.

%% file: Sections/4-AscentioCS.tex
\section{Satellite Ground Segment use-case}

In this case study, we focus exclusively on the ground segment of satellite operations within the context of the Ground Segment as a Service (GSaaS) paradigm. 
The ground segment (GS) plays a critical role in satellite operations, acting as the interface between the space segment and the end users.
The GS handles crucial operations, including telemetry, tracking, command, and the acquisition of mission data. Furthermore, in a GSaaS model, the GS’s cloud-based nature introduces unique opportunities and challenges that warrant targeted analysis. Unlike traditional GSs, which are tightly integrated with proprietary satellite systems, GSaaS decouples the ground infrastructure and offers scalable, on-demand services. This abstraction allows operators to focus on mission objectives while leveraging the flexibility and cost efficiency of shared infrastructure.

The GSaaS under consideration in this case study is a cloud-based solution proposed by Ascentio Technologies S.A., a company with extensive experience in satellite ground infrastructure. Ascentio's GSaaS platform aims to provide satellite operators with seamless access to ground stations (while also allowing the users to interface with their own stations), enabling them to execute uplink and downlink operations, process mission data, and monitor satellite health via cloud-native interfaces. This architecture significantly reduces the overhead associated with building and maintaining dedicated ground infrastructure while offering the scalability required to support multi-mission operations. 

The case study evaluates the safety and security of Ascentio’s GSaaS approach using the proposed AFDT methodology. 
The analysis begins by identifying potential vulnerabilities inherent to cloud-based GSaaS platforms, including risks related to data breaches and service disruptions.
By mapping these vulnerabilities into the AFDT framework, we aim to systematically assess the interactions between security threats, system faults, and implemented defense mechanisms. The resulting AFDT diagram serves as a comprehensive visual representation of risks and mitigations, providing actionable insights for strengthening the platform’s resilience.  

Another critical aspect of the analysis is the identification of MCSs. These cut sets offer valuable information for prioritizing risk mitigation strategies and optimizing resource allocation. A detailed discussion of the analysis of MCSs will be included in a subsequent subsection. 

\begin{figure*}
     \centering
        \includegraphics[width=0.90\textwidth]{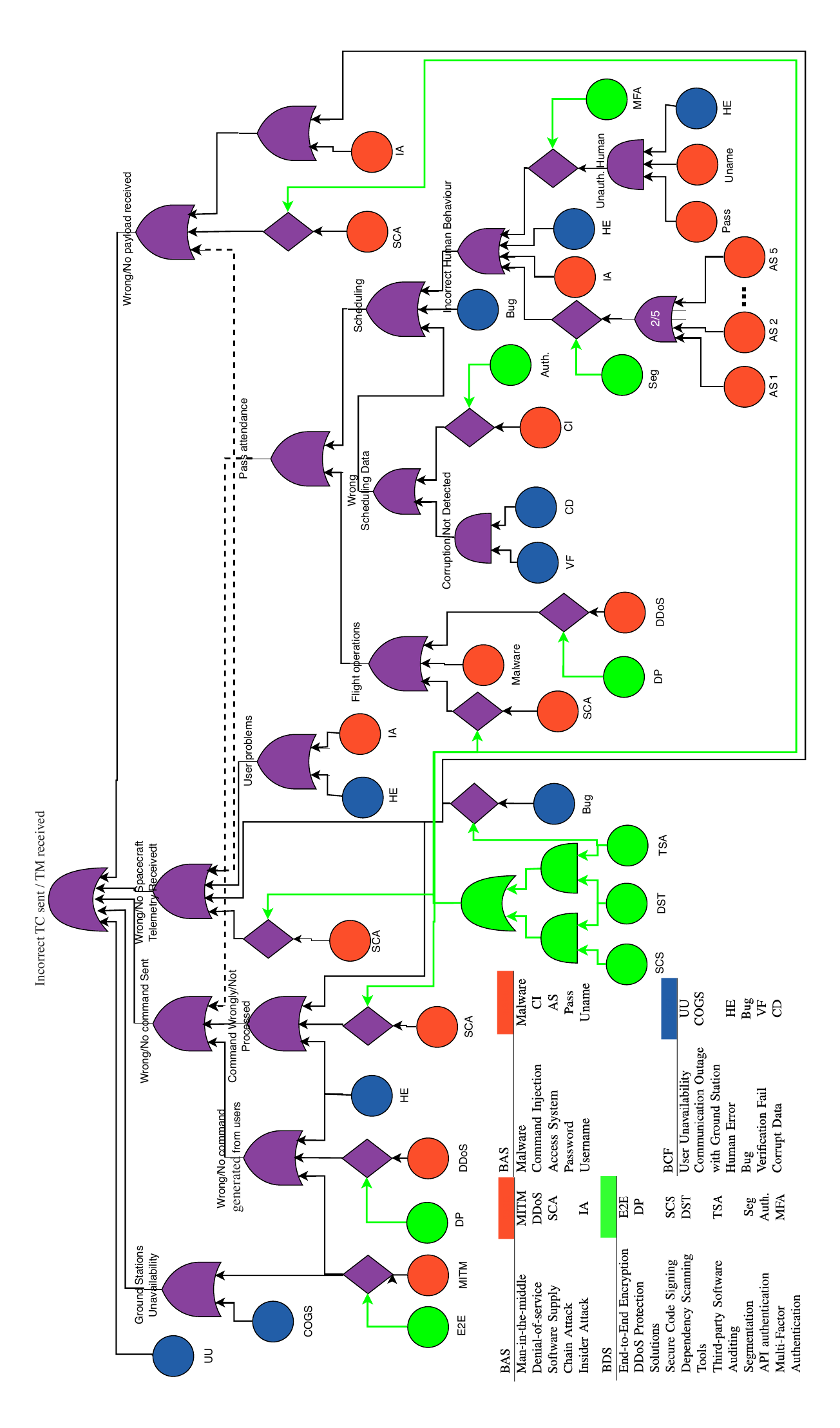}
        \caption{The AFDT of GSaaS}
        \label{afdt}
\end{figure*}

By focusing on the ground segment of Ascentio’s cloud-based GSaaS design, this case study provides a clear and detailed exploration of the paradigm’s implications for safety and security. The AFDT methodology is applied to uncover latent risks, assess defense effectiveness, and propose strategies to enhance the robustness of satellite ground infrastructure in this novel service model. This analysis not only contributes to the ongoing development of Ascentio’s platform but also provides a blueprint for applying AFDTs to other GSaaS implementations as well as other traditional GS infrastructures. Fig. \ref{afdt} depicts the AFDT for the Ascentio’s cloud-based GSaaS.

\subsection{Application of AFDT to a GSaaS}

The AFDT developed for this case study reflects the unique risks and mitigation strategies associated with the cloud-based GS infrastructure, emphasizing its critical functions: telecommand transmission and telemetry reception. 

\paragraph{Structure and Focus.}
The AFDT centers on the top-level event (TLE) defined as the failure to ensure the correct and reliable execution of telecommand and telemetry operations. This TLE branches into intermediate events, key failure pathways specific to the GSaaS environment:
\begin{itemize}
    \item Ground Station Unavailability: \\
    Events leading to a loss of communication between ground segment and satellite, such as hardware malfunctions, connectivity issues, or signal interference.
    \item Faulty Command or Telemetry Data: \\
    Situations where incorrect, incomplete, or corrupted telecommands or telemetry data compromise system integrity.
    \item Human or Procedural Errors: \\
    Mistakes in scheduling, command generation, or manual operations that propagate into larger failures.
\end{itemize}

\paragraph{Iterative Development and Refinement.}
Initial iterations focused on cataloging typical failure modes in GSaaS environments, which were then enriched by analyzing real-world cases of cloud-based system failures. Intermediate events and their dependencies were refined to model how faults and attacks propagate through the GSaaS architecture.
For instance, a DDoS attack on the API endpoints of a virtualized application could lead to service unavailability. This vulnerability is mitigated in the AFDT by modeling defenses such as cloud-based DDoS protection services. 

\subsection{Quantitative analysis of GSaaS}

As specified in Section \ref{background}, MCSs provide information regarding the system’s vulnerability and define the most concise path that causes TLE failure. 
This study presents a qualitative approach to analyzing the system through the identification of MCSs and their most effective defenses.
Examining MCSs gives valuable insights into the system’s reliability, as the system's overall reliability can be augmented by mitigating the attack or failure probability associated with these MCSs. 
By mapping MCSs to their corresponding defenses, the proposed method provides a clear and systematic overview of potential vulnerabilities and mitigation strategies. Even in the absence of quantitative data, this approach proves to be a valuable tool for gaining insights into the interplay between system failures, attacks, and defenses, facilitating better-informed decision-making for enhancing system safety and security.

Table \ref{tab:MCS2} indicates the MCSs related to the GSaaS AFDT of Fig. \ref{afdt}. The table highlights the effectiveness of various defenses in mitigating the MCSs identified for the GSaaS AFDT. Each defense mechanism eliminates the associated MCSs, enhancing the system's overall resilience. For instance, the \emph{E2E} defense is effective against the \emph{MITM} attack, while \emph{DP} mitigates the risk posed by \emph{DDoS}. Similarly, \emph{Seg} is a crucial defense strategy that addresses a wide range of MCSs involving multiple \emph{AS} nodes.

Especially interesting is the defense \emph{TSA}, which protects against both the BCF \emph{Bug} and the BAS \emph{SCA}; and the MCS \{\emph{Pass}, \emph{Uname}, \emph{HE}\}, which consists of both BCFs and BASs. This highlights the importance of analyzing safety, security, and countermeasures in unison.

Certain failures or attacks remain defenseless. This highlights potential vulnerabilities in the system that require further attention to ensure comprehensive risk mitigation. Overall, the analysis of this table underscores the need for a balanced and comprehensive approach to implementing defenses that can effectively mitigate both safety and security risks while identifying and addressing gaps in the current defense mechanisms.

\begin{table}[]
    \centering
    \caption{MCS analysis of the GSaaS AFDT} 
    \begin{tabular}{|c|c|}
    \hline
        MCS & Effective defense(s) \\
        \hline
       \{\textcolor{component}{UU}\} &  \textcolor{defense}{$\emptyset$} \\
       \hline
       \{\textcolor{component}{COGS}\} &  \textcolor{defense}{$\emptyset$} \\
       \hline
       \{\textcolor{attack}{MITM}\} &  \{\textcolor{defense}{E2E}\} \\
       \hline
       \{\textcolor{attack}{DDoS}\} &  \{\textcolor{defense}{DP}\} \\
       \hline
       \{\textcolor{component}{HE}\} &  \textcolor{defense}{$\emptyset$} \\
       \hline
       \{\textcolor{attack}{SCA}\} &  \makecell{\{\textcolor{defense}{SCS, DST}\},\\ \{\textcolor{defense}{DST, TSA}\}} \\
       \hline
       \{\textcolor{attack}{IA}\} &  \textcolor{defense}{$\emptyset$} \\
       \hline
       \{\textcolor{component}{Bug}\} &  \textcolor{defense}{TSA} \\
       \hline
       \{\textcolor{attack}{Malware}\} &  \textcolor{defense}{$\emptyset$} \\
       \hline
       \{\textcolor{component}{VF, CD}\} &  \textcolor{defense}{$\emptyset$} \\
       \hline
       \{\textcolor{attack}{CI}\} &  \{\textcolor{defense}{Auth.}\} \\
       \hline
       \{\textcolor{attack}{AS1, AS2}\} &  \{\textcolor{defense}{Seg}\} \\
       \hline
       \{\textcolor{attack}{AS1, AS3}\} &  \{\textcolor{defense}{Seg}\} \\
       \hline
       \{\textcolor{attack}{AS1, AS4}\} &  \{\textcolor{defense}{Seg}\} \\
       \hline
       \{\textcolor{attack}{AS1, AS5}\} &  \{\textcolor{defense}{Seg}\} \\
       \hline
       \{\textcolor{attack}{AS2, AS3}\} &  \{\textcolor{defense}{Seg}\} \\
       \hline
       \{\textcolor{attack}{AS2, AS4}\} &  \{\textcolor{defense}{Seg}\} \\
       \hline
       \{\textcolor{attack}{AS2, AS5}\} &  \{\textcolor{defense}{Seg}\} \\
       \hline
       \{\textcolor{attack}{AS3, AS4}\} &  \{\textcolor{defense}{Seg}\} \\
       \hline
       \{\textcolor{attack}{AS3, AS5}\} &  \{\textcolor{defense}{Seg}\} \\
       \hline
       \{\textcolor{attack}{AS4, AS5}\} &  \{\textcolor{defense}{Seg}\} \\
       \hline
       \{\textcolor{attack}{Pass, Uname}, \textcolor{component}{HE}\} &  \{\textcolor{defense}{MFA}\} \\
       \hline
    \end{tabular}
    \label{tab:MCS2}
\end{table}

%% file: Sections/5-Conclusion.tex
\section{Conclusion}

This case study underscores the critical value of using the AFDT framework to address the intertwined challenges of safety, security, and defense in complex cyber-physical systems. 
By applying AFDT to the ground segment of a satellite project, we demonstrated how a unified approach identifies vulnerabilities, assesses risks, and guides the development of robust defense strategies, ultimately enhancing the system's overall resilience and reliability.
The presented model aids experts from different fields in uncovering complex dependencies w.r.t. safety and security and analyzing how innovations may affect the security and safety of the intertwined system using MCS analysis.